\documentclass[11pt]{article}
\usepackage{latexsym}
\usepackage{amssymb}
\usepackage[dvips]{graphicx}
\usepackage{epsfig}
\usepackage{rotating}

\usepackage{amsfonts}
\usepackage{amsmath}
\begin{document}

\begin{titlepage}
\begin{flushright}
NITheP-08-12\\
ICMPA-MPA/2008/18\\
\end{flushright}

\begin{center}

{\Large\bf Supersymmetry breaking in\\ noncommutative quantum mechanics}

J Ben Geloun$^{a,b,c,*}$ and F G Scholtz$^{a,\dag}$

$^{a}${\em National Institute for Theoretical Physics}\\
{\em Private Bag X1, Matieland 7602, South Africa}\\
$^{b}${\em International Chair of Mathematical Physics
and Applications}\\
{\em ICMPA--UNESCO Chair 072 B.P. 50  Cotonou, Republic of Benin}\\
$^{c}${\em D\'epartement de Math\'ematiques et Informatique}\\
{\em  Facult\'e des Sciences et Techniques, Universit\'e Cheikh Anta Diop, Senegal}

E-mail:  $^{*}$bengeloun@sun.ac.za,\quad $^{\dag}$fgs@sun.ac.za  

\today

\begin{abstract}
\noindent Supersymmetric quantum mechanics is formulated on a two dimensional noncommutative plane and applied to the supersymmetric harmonic oscillator. We find that the ordinary commutative supersymmetry is partially broken 
and only half of the number of supercharges are conserved. It is argued that this breaking is closely related to the breaking of time reversal symmetry arising from noncommutativity.
\end{abstract}

\end{center}

\end{titlepage}

\setcounter{footnote}{0}

\section{Introduction}
\label{intro}
Ever since the realization that a consistent quantum description of gravity may require a drastic change in our notion of space time at short length scales \cite{dop,seib,wit}, noncommutative geometry, and particularly the formulation of quantum mechanics and quantum field theories on noncommutative spaces \cite{doug} has become a fruitful line of investigation as a possible candidate to replace our current notion of space time \cite{riv}. 

An interesting feature of noncommutative quantum mechanics that emerged from these studies is the breaking of time reversal symmetry in the presence of a non-constant potential \cite{sj,sj7}.  As this clearly lifts some of the degeneracies in the spectrum, it is natural to ask how this may effect other symmetries and in particular supersymmetry.  This is a particular pertinent question in the light of the findings of \cite{seib1} where it was concluded that supersymmetry is half broken in the presence of noncommutativity. 

The simplest setting to discuss supersymmetric quantum mechanics is in the context of the ordinary supersymmetric factorization as discussed in \cite{coop} and references therein. Here our aim is to generalize this supersymmetric factorization to the noncommutative case and to investigate the implications that noncommutativity has for supersymmetry.

As the prime example of a factorizable potential is the harmonic oscillator, this is also the most natural example to which the resulting formalism can be applied.  In contrast to the noncommutative harmonic oscillator on which a considerable body of literature exists (see e.g.\cite{nair,asc,an}), the noncommutative supersymmetric harmonic oscillator only received some attention recently \cite{das,gosh}.  Here we follow a different approach, based on \cite{ioffe}, and rather focus on the issue of supersymmetry breaking, which was not discussed in these papers. 

Not unexpectedly the supersymmetric noncommutative harmonic oscillator Hamiltonian can be diagonalized and the quantum supercharges identified. We find that the noncommutativity partially breaks the ordinary ${\mathcal N}=4$ supersymmetry down to ${\mathcal N}=2$. It is argued that this breaking is directly related to the breaking of time reversal symmetry.

The paper is organized as follows. The next section
is devoted to classical and quantum algebraic
considerations induced by the noncommutativity.
Section 3 develops our main results on noncommutative supersymmetric
factorization which we apply to a solvable case, namely the harmonic
oscillator in two dimensions. 
The paper ends with some discussion in Section 4.

\section{Classical and quantum algebras}
\label{Sect2}

Before studying the supersymmetric version of the noncommutative harmonic oscillator,
this section discuss some algebraic preliminaries pertaining to  
noncommutative coordinate algebras and their representation, thus fixing the notations
of the following sections.  

Noncommutative two dimensional space is defined by the following commutation relation between coordinates
\begin{equation}
[\hat{x}, \hat{y} ]=i\,\theta.
\label{ncb}
\end{equation}
The parameter $\theta$ will be referred to as the noncommutative parameter and has the
dimension of a length squared. More generally, in the $2N$ dimensional case the commutation relations can be brought into a canonical form $[x^i,y^j]=i \Theta^{ij}$, where the antisymmetric tensor $\Theta^{ij}$ has the block diagonal form $\Theta^{ij}={\rm diag}(J^1,J^2,\dots,J^N)$ with $J^{j}$ given by
\begin{eqnarray}
J^j= 
\left(\begin{array}{cc}
0&\theta^j\\
-\theta^j&0
\end{array}\right).
\end{eqnarray}
Thus, for each symplectic pair $(x^j,y^j)$, 
the noncommutative parameter is $\theta^j > 0$, $j=1,2,\ldots, N$.

Introducing the pair of boson annihilation and creation operators $b=(1/\sqrt{2\theta})(\hat{x} + i \hat{y})$ 
and $b^\dagger=(1/\sqrt{2\theta})(\hat{x} - i \hat{y})$, which satisfy
the Heisenberg-Fock algebra $[b,b^\dag]=\,1\!\!1$, noncommutative configuration space is itself a Hilbert space, which we denote by ${\cal H}_c$, isomorphic to boson Fock space ${\cal H}_c = {\rm span}\{|n\rangle, n\in \mathbb{N}\}$, with 
$|n\rangle = (1/\sqrt{n!})(b^\dag)^{n}|0\rangle$.  In the $2N$ dimensional case classical configuration space is simply the $N$ tensorial product of Fock space.  
 
On the quantum level the Hilbert space in which the states of the system are represented, and which we denote by ${\cal H}_q$, is defined to be the space of Hilbert-Schmidt operators on ${\cal H}_c$ \cite{hol}: 
\begin{equation}
\label{qhil}      
\mathcal{H}_q = \left\{ \psi(\hat{x}_1,\hat{x}_2): \psi(\hat{x}_1,\hat{x}_2)\in \mathcal{B}\left(\mathcal{H}_c\right),\;{\rm tr_c}(\psi(\hat{x}_1,\hat{x}_2)^\dagger\psi (\hat{x}_1,\hat{x}_2) < \infty \right\},
\end{equation}
where ${\rm tr}_c$ denotes the trace over ${\cal H}_c$ and
$\mathcal{B}\left(\mathcal{H}_c\right)$ is the set of bounded operators
on $\mathcal{H}_c$.

Next, we seek a representation on ${\cal H}_q$ of the noncommutative Heisenberg algebra 
\begin{eqnarray} 
&&\begin{array}{rcl}
&&[\hat{X},  \hat{Y} ] = i\theta, \\
&&[\hat{X}, \hat{P}_{X} ] = i\hbar=
[\hat{Y}, \hat{P}_{Y} ],  \\
&&[\hat{P}_{X}, \hat{P}_{Y} ]= 0.
\end{array}
\label{qbalg}
\end{eqnarray} 
Henceforth capital letters are reserved to refer to quantum operators acting on ${\cal H}_q$ in order to distinguish them from operators acting on noncommutative configuration space ${\cal H}_c$. It is easily verified that a representation is provided by the following  
\begin{eqnarray} 
&&\hat{X} \psi= \hat{x} \psi, \quad
\hat{Y} \psi =  \hat{y}\psi , \quad 
\hat{P}_{X} \psi =\frac{\hbar}{\theta} [\hat{y},\psi], \quad
\hat{P}_{Y} \psi= -\frac{\hbar}{\theta} [\hat{x},\,\psi].
\label{brep}
\end{eqnarray}
Furthermore, it can also be shown that these operators
are self-adjoint with respect to the quantum Hilbert space inner product 
$(\phi|\psi)={\rm tr_c} (\phi^\dag \psi)$, which makes this a unitary representation.

\section{Supersymmetric noncommutative harmonic oscillator}

With the formal structure of the noncommutative quantum theory settled, 
we proceed to introduce the concept of supersymmetric
factorization in a noncommutative space after which we apply the resulting formalism to the noncommutative harmonic oscillator.

\subsection{Noncommutative supersymmetric factorization} 

Supersymmetric factorization in commutative quantum mechanics and in two or more dimension has recently been considered in \cite{ioffe}.  Here we generalize this approach to a noncommutative quantum system focusing on two dimensions.  

We consider the following noncommutative Hamiltonian
\begin{eqnarray}
H= \frac{1}{2m} ( \hat P_X^2 +  \hat P_Y^2 )  + V_1(\hat X, \hat Y),
\end{eqnarray}
where $V_1(\hat x, \hat y)$, considered as an operator acting on configuration space, is Hermitian, which is the analogue of a real potential in commutative space.  The factorized quantum Hamiltonian assumes the general form
\begin{eqnarray}
H_1\, = \hbar \omega\, (\,  B^\dag_X B_X \,+\,  B^\dag_Y  B_Y\,), 
\label{facto1}
\end{eqnarray}
where the dimensionless operators $B_X, B^\dag_X, B_Y, B^\dag_Y$ are defined as
\begin{eqnarray}
&&B_X = \frac{1}{\sqrt{\hbar\omega}}(\frac{i}{\sqrt{2m}}\hat P_X + W_X(\hat X, \hat Y)),  \qquad
B^\dag_X =\frac{1}{\sqrt{\hbar\omega}}\, ( - \frac{i}{\sqrt{2m}} \hat P_X + W_X(\hat X, \hat Y)),\label{bx}\\
&&B_Y = \frac{1}{\sqrt{\hbar\omega}}\, (\frac{i}{\sqrt{2m}}\hat P_Y + W_Y(\hat X, \hat Y)), \qquad
B^\dag_Y = \frac{1}{\sqrt{\hbar\omega}}\, (- \frac{i}{\sqrt{2m}}\hat P_Y + W_X(\hat X, \hat Y)),
\label{bydag}
\end{eqnarray}
and the Hermitian superpotentials $W^\dag_X(\hat X, \hat Y) = W_X(\hat X, \hat Y)$ and 
$W^\dag_Y(\hat X, \hat Y) = W_Y(\hat X, \hat Y)$ are yet to be specified.
From (\ref{facto1}) one obtains the noncommutative version
of the Riccati equation in two dimensions
\begin{eqnarray}
V_1(\hat X, \hat Y) =
\frac{i}{\sqrt{2m}} \left( (\hat P_X W_X)(\hat X, \hat Y) + (\hat P_Y W_Y)(\hat X, \hat Y)\right)
+ (W_X(\hat X, \hat Y))^2 + (W_Y(\hat X, \hat Y))^2, 
\end{eqnarray}  
which is closely related to the so called algebraic (matrix) Riccati equation.
One notices that the two superpotentials $W_X$ and $W_Y$ are coupled.
The ground state operator $\Psi_{00}$ should be annihilated
by both of the annihilators $B_X$ and $B_Y$, but in contrast to the commutative case the solutions $W_X$ and $W_Y$ 
of the above equation cannot be written explicitly in terms of the ground state, but 
are given by an algebraic system 
\begin{eqnarray}
&&\frac{i\hbar}{\sqrt{2m}\theta} [\hat Y,\Psi_{00} ]+ W_X(\hat X, \hat Y)\Psi_{00}=0,\\
&&\frac{-i\hbar}{\sqrt{2m}\theta} [\hat X,\Psi_{00} ] + W_Y(\hat X, \hat Y)\Psi_{00}=0.
\end{eqnarray} 
As is customarily the case in one dimensional supersymmetry quantum mechanics, 
a different Hamiltonian can be obtained by reversing the order of the operators.
Here, apart from $H_1$, we get mixed types of operators \cite{ioffe}
\begin{eqnarray}
&&H'_{11} = \hbar \omega\, (\,  B_X B^\dag_X\,+\,B^\dag_Y  B_Y \,),
\qquad H'_{22}= \hbar \omega\, (\, B^\dag_X B_X \,+\,  B_Y B^\dag_Y\,),\\
&&H'_{12} = \hbar \omega\, (\,  B_Y B^\dag_X\,-\,B^\dag_X  B_Y \,),
\qquad H'_{21}= \hbar \omega\, (\, B_X B^\dag_Y  \,-\,B^\dag_Y B_X \,),\\
&&H_2\, = \hbar \omega\, (\,  B_X B^\dag_X\,+\,  B_Y B^\dag_Y\,),
\label{facto2}
\end{eqnarray}
with ${H'}^\dag_{12} =  H'_{21}$.
The corresponding superpartner potentials can also be derived
\begin{eqnarray}
&V'_1(\hat X, \hat Y) =
\frac{-i}{\sqrt{2m}} \left[ (\hat P_X W_X)(\hat X, \hat Y) - (\hat P_Y W_Y)(\hat X, \hat Y)\right]
+ (W_X(\hat X, \hat Y))^2 + (W_Y(\hat X, \hat Y))^2,\\ 
&V'_2(\hat X, \hat Y) =
 \frac{i}{\sqrt{2m}} \left[(\hat P_X W_X)(\hat X, \hat Y) -  (\hat P_Y W_Y)(\hat X, \hat Y)\right]
+ (W_X(\hat X, \hat Y))^2 + (W_Y(\hat X, \hat Y))^2,\\
&V_2(\hat X, \hat Y) =
\frac{-i}{\sqrt{2m}} \left[ (\hat P_X W_X)(\hat X, \hat Y) +  (\hat P_Y W_Y)(\hat X, \hat Y)\right]
+ (W_X(\hat X, \hat Y))^2 + (W_Y(\hat X, \hat Y))^2.
\end{eqnarray}
Here $V'_1,V'_2$ and $V_2$ are, respectively, associated with $H'_{11},H'_{22}$ and $H_2$.  
We emphasize that this factorization method is more general than the
direct noncommutative extension of the two dimensional commutative supersymmetric formulation 
of \cite{ioffe}. Indeed, in the latter an unique superpotential $W(\hat X, \hat Y)$ 
is required and raising and lowering operators are defined by substituting in 
(\ref{bx})-(\ref{bydag}) $W_X(\hat X, \hat Y)=(\hat  P_X W)(\hat X, \hat Y)$ and 
$W_Y(\hat X, \hat Y)=(\hat  P_Y W)(\hat X, \hat Y)$. 
The two dimensional Riccati equation in the variables $(\hat P_X W, \hat P_Y W)$  can then be written in terms of a generalized noncommutative gradient of $\hat P_{X,Y}W$, namely 
\begin{eqnarray}
V_1(\hat X, \hat Y) =
\frac{i}{\sqrt{2m}} \left( \hat P^2_X W + \hat P^2_Y W\right)(\hat X, \hat Y)
+ ((\hat P_X W)^2 + (\hat P_Y W)^2 )(\hat X, \hat Y).
\end{eqnarray} 
The natural question of the relation between energy eigenvalues and eigenfunctions
of the different Hamiltonian $H_1, H'_1, H'_2$ and $H_2$ can now be considered, i.e., 
the supercharge formulation should be investigated.
The following relations hold
\begin{eqnarray}
&& H_1 B^\dag_X = B^\dag_X H'_{11} +B^\dag_Y H'_{21} + [B^\dag_Y, B^\dag_X] B_Y, 
\label{su1}\\
&& H_1 B^\dag_Y = B^\dag_X H'_{12} +B^\dag_Y H'_{22} - [B^\dag_Y, B^\dag_X] B_X,\\ 
&& B^\dag_X H_2 =  H'_{22} B^\dag_X -  H'_{21}B^\dag_Y - B_Y[B^\dag_Y,B^\dag_X],\\
&& B^\dag_Y H_2 = - H'_{12} B^\dag_X +  H'_{11} B^\dag_Y + B_X[B^\dag_Y,B^\dag_X].
\label{su4}  
\end{eqnarray}
Other types of relations can be obtained by taking the adjoint of 
these identities. Note that the last terms in (\ref{su1})-(\ref{su4}) 
involve the commutator $[B^\dag_Y, B^\dag_X]$, which
cannot vanish without further assumptions. It turns out that,
choosing  $W_X(\hat X, \hat Y)=(\hat  P_X W)(\hat X, \hat Y)$ and 
$W_Y(\hat X, \hat Y)=(\hat  P_Y W)(\hat X, \hat Y)$, this commutator
reduces to
\begin{eqnarray} 
[B^\dag_Y, B^\dag_X] =[\hat P_Y W, \hat P_X W],
\end{eqnarray}
and that in the commutative limit one recovers the two dimensional
supersymmetry as discussed in \cite{ioffe}. This underlines also
the plausible scenario that noncommutativity could break the supersymmetry.
Another issue that we shall not pursue here, but that could be of major interest, is the notion of noncommutative shape invariance, solvable potentials and Hamiltonian hierarchy \cite{coop}. 
As these concepts rest on algebraic commutation relations and the noncommutativity involves the adjoint action, it should be possible to investigate these concepts in the noncommutative setting.
However, due to the dimension greater than one, these notions are not even yet well understood in the commutative case \cite{pan}. 

\subsection{Application to the noncommutative harmonic oscillator}

As an application of the above formalism, let us consider the noncommutative harmonic oscillator.
The quantum Hamiltonian defining the motion of a non-relativistic 
particle of mass $m$ confined in a harmonic well with
frequency $\omega$ within a noncommutative plane can be written up to a constant\footnote{This constant $-\hbar \omega$ has to be related to the unbroken supersymmetry as appears hereafter.}  as
\begin{eqnarray}
H_1&=& \frac{1}{2m}(\hat{P}^{\,2}_{X} +\hat{P}^{\,2}_{Y}) + \frac{1}{2}m \omega^2 (\hat{X}^2 + \hat{Y}^2) - \hbar \omega.
\end{eqnarray}
The following superpotentials 
\begin{eqnarray}
W_X(\hat X, \hat Y)= \sqrt{\frac{m\omega}{2 \hbar}} \hat X ,\qquad
W_Y(\hat X, \hat Y)= \sqrt{\frac{m\omega}{2 \hbar}} \hat Y,
\end{eqnarray}
allow the following definition of the ladder operators:
\begin{eqnarray}
&& B_X = \sqrt{\frac{m\omega}{2 \hbar}}\left(\hat X +\frac{i}{m\omega} \hat P_X\right),\quad\;
B^\dag_X = \sqrt{\frac{m\omega}{2 \hbar}}\left(\hat X - \frac{i}{m\omega} \hat P_X\right),\\
&& B_Y = \sqrt{\frac{m\omega}{2 \hbar}}\left(\hat Y +\frac{i}{m\omega} \hat P_Y\right),\quad\;
B^\dag_Y = \sqrt{\frac{m\omega}{2 \hbar}}\left(\hat Y -\frac{i}{m\omega} \hat P_Y\right).
\end{eqnarray}
These operators admit the following factorization of the Hamiltonian $H_1$:
\begin{eqnarray}
H_1= \hbar \omega \left( B_X^\dag B_X + B^\dag_Y B_Y \right).
\label{facto3}
\end{eqnarray}
The associated Hamiltonians are easily computed
\begin{eqnarray}
H_2= H_1 + 2\hbar \omega \mathbb{I},\quad
H'_{11} = H_1 +\hbar\omega = H'_{22}, \quad H'_{12} = -\frac{i m \omega^2 \theta}{2} = - H'_{21}.
\end{eqnarray}
The noncommutative supersymmetric factorization can therefore be carried out exactly for the noncommutative harmonic
oscillator. However, the quantum supercharges cannot be identified with 
these operators. Indeed, we can check that
$[B_X^\dag,B_Y^\dag]\propto \theta$ implies that (\ref{su1})-(\ref{su4})
do not define a superalgebra. It is only through a re-factorization of 
the Hamiltonian with respect to decoupled operators that the superalgebra
emerges.

To proceed, we write $H_1$ (\ref{facto3}) in the matrix form 
\begin{eqnarray}
H= \hbar\omega\; B^\dag\,B,\quad
\;\; B =(B_X, B_Y)^t, 
\label{hamat}
\end{eqnarray}
where symbol $t$ denotes the transpose operation. The purpose of this rewriting is the factorization of the Hamiltonian in terms of diagonal bosonic operators:
\begin{eqnarray}
&&H= \hbar\omega\;
A^\dag\, D\, A ,\quad A=(A_+,A_-)^t, 
\label{hfact}
\end{eqnarray}
where $D$ is some diagonal positive matrix and 
$(A_\pm,A_\pm^\dag)$ 
satisfy decoupled and diagonal bosonic commutation relations, 
\begin{eqnarray} 
[A_\pm,A^\dag_\pm] = \mathbb{I}.
\label{nwqbalg}
\end{eqnarray}
We introduce the vectors $A^+= (A^\dag_+, A^\dag_-)^t$ and $B^+=(B^\dag_X, B^\dag_Y)^t$, for which 
$(A^+)^t = A^\dag$, as well as a linear transformation $S$ relating them
\begin{eqnarray}
A = S\, B, \quad A^+ = S^* \,B^+. 
\label{rela2}
 \end{eqnarray}
Defining the matrix $\mathfrak{g}$ with elements 
\begin{eqnarray}
\mathfrak{g}_{kl} =  [ B_k, B^+_l], \quad k,l=1,2,\quad B_1 := B_X, \;\;\, B_2 := B_Y,
\end{eqnarray}
it is simple to verify that $\mathfrak{g}$ is Hermitian. Let us denote its eigenvectors by $u_i$, $i=1,2$.
From the commutation relations
\begin{eqnarray}
[A_i, A^+_j] =  \delta_{ij},\;\; i,j=1,2,
\end{eqnarray}
one derives the identity
\begin{eqnarray}
S \, \mathfrak{g}\, S^\dag = \mathbb{I}_{2}.
\label{key}
\end{eqnarray} 
The Hamiltonian diagonalization is now immediate with the following choice of the matrix 
$S^\dag= (u_1, u_2)$.  Noting from (\ref{key}) that $(S^\dag)^{-1}=  S\,\mathfrak{g}$,
we get from (\ref{rela2}), when inserted into (\ref{hamat}), the following result
\begin{eqnarray}
H= \hbar\omega \; (A^+)^t \,S\, \mathfrak{g}^2\, S^\dag \, A.
\end{eqnarray} 
It simply remains to apply $\mathfrak{g}$ twice
on its eigenstates to obtain the diagonal form
of the Hamiltonian. The eigenvectors
have not yet been normalized. The normalization conditions are fixed from the requirement that for $ i=1,2$, $(u_i)^\dag u_i =1/ |\lambda_i|$, where $\lambda_i$ is the eigenvalue associated with $u_i$. Thus, in terms of these boson operators the Hamiltonian can be expressed as
\begin{equation}
H= \hbar\omega \left( |\lambda_+| A^\dag_+ A_+ +  |\lambda_-| A^\dag_- A_-\right),
\label{diagham}
\end{equation} 
where $|\lambda_\pm|$ are the absolute values of the eigenvalues of the matrix
\begin{equation} 
\mathfrak{g}=
\left(\begin{array}{cc}
1 &\mathbf{a}_\theta\\
-\mathbf{a}_\theta&1
\end{array}\right) \,.
\label{gg}
\end{equation}
Here we have introduced the parameter $\mathbf{a}_\theta=\frac{1}{2\hbar} m \omega\theta$. 
The computation of these eigenvalues is easily performed with the result
\begin{equation} 
|\lambda_{+}| = 1+ \mathbf{a}_\theta, \qquad
|\lambda_{-}| = |1- \mathbf{a}_\theta|,
\end{equation}
which diagonalizes the Hamiltonian (\ref{hamat}).
This operator is nothing but the Hamiltonian describing an harmonic oscillator with frequency encoding the noncommutative parameters $(\theta,\hbar)$. In terms of $B$ these bosonic operators can be expressed as
\begin{eqnarray}
&&A_+ = \frac{1}{c_+}(- iB_X +B_Y), \qquad A_- = \frac{1}{c_-}(iB_X +B_Y),\cr
&&c_+ = \sqrt{2 (1+\mathbf{a}_\theta )},\quad  
c_- = \sqrt{2 |1-\mathbf{a}_\theta |},
\label{adiag}
\end{eqnarray}
while $A_\pm^\dag$ are obtained by the adjoint. Finally, we mention that this factorization still works for a model with broken rotational symmetry resulting from different frequencies
$(\omega_X, \omega_Y)$ for the noncommutative directions. In this case, 
one considers, without loss of generality, scaled operators in one direction,
for instance $\hat Y$, namely,  
$\widetilde {B}_Y= \sqrt{(\omega_Y/\omega_Y)} B_Y$, and proceeds in the same
way as before. One arrives at the following factorized operator
\begin{eqnarray}
& H'= \hbar \left( |\lambda'_+| A'^{\,\dag}_+ A'_+ +  |\lambda'_-| A'^{\,\dag}_- A'_-\right),\\
&|\lambda'_\pm|= \left|\frac{\omega_X +\omega_Y}{2} \pm \frac{1}{2}\sqrt{(\omega_X -\omega_Y)^2 + (\frac{m\omega_X\omega_Y \theta}{\hbar})^2}\right|,\nonumber
\end{eqnarray}
which, of course reduces to (\ref{diagham}) when $\omega_X=\omega_Y= \omega$.

At this point, we are able to identify the quantum supercharges.
Fixing the constant parameters $\kappa_\pm = \sqrt{\hbar\omega|\lambda_\pm|}$,
the following operator
\begin{eqnarray}
Q=\left(\begin{array}{cccc}
0&0&0&0\\
\kappa_+ A_+&0&0&0\\
\kappa_- A_-&0&0&0\\
0&\kappa_- A_-&-\kappa_+ A_+&0
\end{array}\right)
\end{eqnarray}
and its adjoint $Q^\dag$ satisfy a ${\mathcal N} =2$ superalgebra
such that 
\begin{eqnarray}
\{Q,Q^\dag\} ={\mathcal H},\quad [Q,{\mathcal H}]=0=[Q^\dag,{\mathcal H}],
\end{eqnarray}
with the $4\times 4$ matrix Hamiltonian given by
\begin{eqnarray}
&&{\mathcal H}=\left(\begin{array}{cccc}
{\mathcal H}^1_+ + {\mathcal H}^1_- &0&0&0\\
0& {\mathcal H}^2_+ +{\mathcal H}^1_- & 0&0\\
0& 0&{\mathcal H}^2_- +{\mathcal H}^1_+ &0\\
0&0&0&{\mathcal H}^2_+ +{\mathcal H}^2_-\end{array}\right),\\
&&{\mathcal H}^1_\pm = \hbar\omega |\lambda_\pm| A^\dag_\pm A_\pm, \quad
{\mathcal H}^2_\pm = \hbar\omega  |\lambda_\pm| A_\pm A^\dag_\pm.
\end{eqnarray}
We get the following relations between the Hamiltonians 
\begin{eqnarray}
&& {\mathcal H}_{11}\, (\kappa_+ A^\dag_+)\, =\,(\kappa_+ A^\dag_+)\, 
{\mathcal H}_{22},\qquad
{\mathcal H}_{11}\, (\kappa_+ A^\dag_-)\, =\, (\kappa_+ A^\dag_-) {\mathcal H}_{33} , \label{su11}\\ 
&& (\kappa_+ A^\dag_+)\, {\mathcal H}_{44} =  {\mathcal H}_{33}\,  (\kappa_+ A^\dag_+), \qquad
(\kappa_+ A^\dag_-) \, {\mathcal H}_{44} =  {\mathcal H}_{22}\,  (\kappa_+ A^\dag_+),
\label{su41}  \\
&&{\mathcal H}_{11} = {\mathcal H}^1_+ + {\mathcal H}^1_-,\;\;
{\mathcal H}_{22} = {\mathcal H}^2_+ + {\mathcal H}^1_-,\;\;
{\mathcal H}_{33} = {\mathcal H}^1_+ + {\mathcal H}^2_-,\;\;
{\mathcal H}_{44} = {\mathcal H}^2_+ + {\mathcal H}^2_-,
\end{eqnarray}
which are to be compared with (\ref{su1})-(\ref{su4}). From this one can obtain the 
correct relations between the operator eigenfunctions of these Hamiltonians. 

Each of the matrix operators $Q$ and $Q^\dag$ can indeed be decomposed
into four elementary matrices labeled by their entries and each of these
``sub-operators'' represent a symmetry of the Hamiltonian. In that language, 
we should say that the model is ${\mathcal N}=8$ supersymmetric.
Nevertheless, as a matter of compact notation we keep to the operators $Q$ and $Q^\dag$, which generate a 
${\mathcal N}=2$ supersymmetry. It should be emphasized that in the commutative limit $\theta\to 0$ the operator 
\begin{eqnarray}
Q'=\left(\begin{array}{cccc}
0&0&0&0\\
\kappa_- A_-&0&0&0\\
\kappa_+ A_+&0&0&0\\
0&\kappa_+ A_+&-\kappa_- A_-&0
\end{array}\right),
\end{eqnarray}
and its adjoint $Q'^\dag$ also provide a set of symmetries for the
Hamiltonian. In the presence of noncommutative geometry, i.e., when the noncommutative parameter $\theta\ne 0$, the total supersymmetry is partially broken and the number of supercharges
decreases from ${\mathcal N}=4$ to ${\mathcal N}=2$ 
(actually from ${\mathcal N}=16$ to ${\mathcal N}=8$).
Finally, let us mention that this partial supersymmetry breaking is
in agreement with the supersymmetry reduction ${\cal N}=1 \to 1/2$ 
as pointed out by Seiberg in the context of
noncommutative superfield theory in four dimensional noncommutative spacetime \cite{seib}.

Let us now investigate the relationship between this partial supersymmetry
breaking and time reversal asymmetry arising from noncommutativity.
We start by calculating the total angular momentum
operator. Given the tensor product supercharges
\begin{eqnarray}
&&Q_1= \kappa_+ A_+\,\, \sigma_- \otimes \sigma_3 ,\qquad
Q^\dag_1= \kappa_+ A^\dag_+\,\, \sigma_+ \otimes \sigma_3, \\
&&Q_2= \kappa_- A_-\,\, \mathbb{I}_2 \otimes \sigma_-,\qquad
Q^\dag_2= \kappa_- A^\dag_-\,\, \mathbb{I}_2 \otimes \sigma_+,
\end{eqnarray}
we can write 
\begin{eqnarray}
Q = Q_1 + Q_2,\quad Q^\dag = Q^\dag_1 + Q^\dag_2.
\end{eqnarray}
The Hamiltonian can then be written as
\begin{eqnarray}
{\cal H}&=& \{Q_1,Q^\dag_1 \} + \{ Q_2, Q^\dag_2\}
= {\cal H}'\,\,\mathbb{I}_2 \otimes \mathbb{I}_2  - \frac{\kappa^2_+}{2} \sigma_3\otimes  \mathbb{I}_2    - \frac{\kappa^2_-}{2} \mathbb{I}_2 \otimes \sigma_3,\cr
{\cal H}' &=& \kappa^2_+ A^\dag_+ A_+\,+ \,\kappa^2_- A^\dag_- A_-\,\,
+ \frac{\kappa^2_+}{2} + \frac{\kappa^2_-}{2}.
\end{eqnarray}
The system therefore admits a total spin operator as
\begin{eqnarray}
S=\frac{\hbar}{2} \left( - \mathbb{I}_2 \otimes \sigma_3 + \sigma_3\otimes  \mathbb{I}_2 \right).
\end{eqnarray}
The following identities can be deduced
\begin{eqnarray}
[S\,,\, \sigma_- \otimes \sigma_3]\, =\,  - \hbar\, \sigma_- \otimes \sigma_3 \qquad
[S\,,\,\mathbb{I}_2 \otimes \sigma_-]\,=\,  \hbar\, \mathbb{I}_2 \otimes \sigma_- .
\label{spin}
\end{eqnarray}
The noncommutative orbital angular momentum operator $L_z$ is given by \cite{sj7}
\begin{eqnarray} 
L_z= \hat X \hat P_Y - \hat Y \hat P_X + \frac{\theta}{2\hbar} 
(\hat P^2_X + \hat P^2_Y)
\end{eqnarray}
and satisfies the algebra
$[L_z, \hat X] = i \hbar \hat Y$ and $[L_z, \hat Y] =  -i\hbar \hat X$. Using the inverse relations (\ref{adiag}), the following commutation relations can be obtained
\begin{eqnarray}
[L_z, A_\pm] =\pm\, \hbar\, A_\pm, \qquad
[L_z, A^\dag_\pm] =\mp\,\hbar\, A^\dag_\pm.
\label{angu}
\end{eqnarray}
Finally, the total angular momentum operator is  
\begin{eqnarray} 
J = L_z \, \mathbb{I}_4  + S. 
\end{eqnarray}
Using (\ref{spin}) and (\ref{angu}), the supercharge $Q$ transforms as follows 
\begin{eqnarray}
[J, Q]  = 
\kappa_+([L_z,A_+] - \hbar A_+ ) \,\, \sigma_- \otimes \sigma_3 + 
\kappa_-([L_z,A_-] + \hbar A_- )\,\, \mathbb{I}_2 \otimes \sigma_-=0.
\end{eqnarray}
Therefore, the supercharge $Q$
and the angular momentum $L_z$ are commuting objects.
On the other hand, one can immediately check that $[L_z, Q']\neq 0$ 
since by switching the operators $A_+ \to A_-$, this commutator becomes nontrivial. 
As breaking of time reversal symmetry lifts the degeneracy between states
with angular momentum $+m$ and $-m$, it is clear that only the supercharge
$Q$ that does not change the angular momentum can survive in the
noncommutative limit. Indeed, it is not difficult to check that
this is exactly half of the supercharges in the commutative case.

\section{Conclusion}
\label{Sect3}

We have extended the fundamental notions of supersymmetric factorization to
two dimensional noncommutative quantum systems. The new approach was then applied 
to factorize the noncommutative harmonic oscillator. 
The diagonalization of the quantum Hamiltonian operator has been successfully performed and, consequently, we determined the quantum supercharges. 
It turns out that the supersymmetry is partially broken and the number of supercharges decreases in the presence of noncommutativity. 
Furthermore, we have discussed how this supersymmetry breaking is related to time reversal symmetry breaking. 
Finally, in the commutative limit the usual properties are recovered.

\section*{Acknowledgments}
This work was supported under a grant of the  National Research Foundation of South Africa.

\end{document}